\documentclass[preprint2]{aastex}

\slugcomment{To appear in ApJ}

\shorttitle{Rb-rich MC AGB stars}
\shortauthors{Garc\'{\i}a-Hern\'andez et al.}

\begin{document}


\title{Rb-rich Asymptotic Giant Branch stars in the Magellanic Clouds}


\author{D. A. Garc\'\i a-Hern\'andez\altaffilmark{1}, A.
Manchado\altaffilmark{1,2}, D. L. Lambert\altaffilmark{3}, B.
Plez\altaffilmark{4}, P.  Garc\'\i a-Lario\altaffilmark{5}, F.
D'Antona\altaffilmark{6}, M. Lugaro\altaffilmark{7}, A. I.
Karakas\altaffilmark{8} and M. A. van Raai\altaffilmark{9}}


\altaffiltext{1}{Instituto de Astrof\'{\i}sica de Canarias, C/ Via L\'actea
s/n, 38200 La Laguna, Spain; agarcia@iac.es}
\altaffiltext{2}{Consejo Superior de Investigaciones Cient\'{\i}ficas, Spain}
\altaffiltext{3}{W. J. McDonald Observatory. The University of Texas at
Austin. 1 University Station, C1400. Austin, TX 78712$-$0259, USA; dll@astro.as.utexas.edu}
\altaffiltext{4}{GRAAL, Universit\'e Montpellier 2, CNRS, Montpellier, France}
\altaffiltext{5}{Herschel Science Centre. European Space Astronomy Centre,
Research and Scientific Support Department of ESA. Villafranca del Castillo,
P.O. Box 50727. E-28080 Madrid. Spain;  Pedro.Garcia-Lario@sciops.esa.int}
\altaffiltext{6}{INAF-Osservatorio Astronomico di Roma}
\altaffiltext{7}{Centre for Stellar and Planetary Astrophysics, Monash University, Clayton 3800, Victoria, Australia}
\altaffiltext{8}{Research School of Astronomy and Astrophysics, Mt Stromlo Observatory, Weston Creek ACT 2611, Australia}
\altaffiltext{9}{Sterrenkundig Instituut, University of Utrecht, Postbus 80000, 3508 TA Utrecht, The Netherlands}


\begin{abstract}
We present high-resolution (R$\sim$60,000) optical spectra of a carefully
selected sample of heavily obscured and presumably massive O-rich Asymptotic
Giant Branch (AGB) stars in the Magellanic Clouds (MCs). We report the discovery
of strong Rb\,{\sc i} lines at 7800 \AA~in four Rb-rich LMC stars at
luminosities equal to or greater than the standard adopted luminosity limit for
AGB stars (M$_{bol}$$\sim$$-$7.1), confirming that ``Hot Bottom Burning" (HBB)
may produce a flux excess in the more massive AGB stars. In the SMC sample, just
one of the five stars with M$_{bol}$$<$$-$7.1 was detected in Rb; the other
stars may be massive red supergiants. The Rb-rich LMC AGB stars might have
stellar masses of at least $\sim$6$-$7 M$_{\odot}$. Our abundance analysis show
that these Rb-rich stars are extremely enriched in Rb by up to
10$^{3}$$-$10$^{5}$ times solar but seem to have only mild Zr enhancements. The
high Rb/Zr ratios, if real, represent a severe problem for the $s$-process, even
if the $^{22}$Ne source is operational as expected for massive AGB stars; it is
not possible to synthesize copious amounts of Rb without also overproducing Zr.
The solution to the problem may lie with an incomplete present understanding of
the atmospheres of luminous AGB stars.
\end{abstract}





\section{Introduction}

The Magellanic Clouds (hereafter MCs) provide a unique opportunity to study the
evolution and nucleosynthesis of low- and intermediate-mass stars (0.8 $<$ M $<$
8 M$_{\odot}$) in low metallicity environments without the uncertainty on
distance, and hence luminosity that hinders similar studies in our Galaxy. Low-
and intermediate-mass stars experience thermal pulses on the Asymptotic
Giant Branch (AGB) and end their lives with a phase of strong mass loss (see
e.g., Herwig 2005). The thermal pulses driven by He-burning
produce $^{12}$C which is mixed to the stellar surface via the ``3$^{rd}$
dredge-up" following a pulse. However, in the case of the more massive AGB stars
(M$>$4 M$_{\odot}$), the base of the convective envelope is predicted to
experience H-burning by Hot Bottom Burning (HBB, e.g., Mazzitelli et al. 1999),
so these stars remain O-rich despite the dredge-up. HBB models predict the
production of $^{13}$C and $^{14}$N through the CN cycle as well as $^{7}$Li
enhancements. Occurrence of HBB in massive AGB stars is supported by previous
studies on visually bright MC AGB stars (Plez et al. 1993; Smith et al. 1995).
In our own Galaxy, we have recently identified a small group of obscured but
infrared-bright stars showing OH maser emission, the OH/IR stars (e.g., Wood et
al. 1992), as very massive (4$-$8 M$_{\odot}$) O-rich AGB stars experiencing
HBB, as indicated by their strong Li overabundances (Garc\'{\i}a-Hern\'andez et
al. 2007). 

Theoretical models predict also the presence of $s$-process elements such as Rb,
Zr, Sr, etc. at the stellar surface as a consequence of the ``3$^{rd}$
dredge-up" episodes (Busso et al. 1999). According to recent understanding,
$^{13}$C($\alpha$,n)$^{16}$O is the preferred neutron source in the He-shell for
masses around 1$-$4 M$_{\odot}$, while for more massive stars neutrons are
mainly released by $^{22}$Ne($\alpha$,n)$^{25}$Mg. This is because activation of
$^{22}$Ne requires the higher temperatures (T $>$ 3 x 10$^{8}$ K) reached during
thermal pulses in more massive AGB stars (Lugaro \& van Raai 2008). The relative
abundance of Rb to other nearby s-elements such as Zr (i.e., the Rb/Zr ratio) is
sensitive to the neutron density owing to branchings in the s-process path at
$^{85}$Kr and $^{86}$Rb (Lambert et al. 1995; Abia et al. 2001; van Raai
et al. 2008). Since the $^{22}$Ne neutron source produces much higher neutron
densities than the $^{13}$C neutron source, the Rb/Zr ratio is a discriminant of
the operation of the $^{13}$C versus the $^{22}$Ne neutron source and, as such,
a good indicator of the progenitor stellar mass in AGB stars.

Interestingly, we discovered  strong Rb overabundances (up to 10$-$100 times
solar) with apparently only mild Zr enhancements in massive galactic O-rich
AGB stars (Garc\'{\i}a-Hern\'andez et al. 2006, 2007). This work provided the
first observational suggestion that $^{22}$Ne is the dominant neutron source in
(presumably) massive AGB stars. Rb was not found to be overabundant in the few
unobscured O-rich massive AGBs previously studied in the SMC (Plez et al. 1993).
With our strategy that was successful in finding massive AGB stars in the
Galaxy, we extended optical spectroscopy to obscured and presumably massive MC
O-rich AGB stars. Here, we report for the first detections of massive Rb-rich
AGB stars in the MCs. 

\section{Optical observations}

A sample of 24 obscured O-rich AGBs in the MCs was carefully selected from
the literature. Most of the known MC OH/IR stars (Wood et al. 1992; Marshall et
al. 2004 and references therein) were included in our sample, as they are, in
principle, the most massive and extreme AGBs known in the Clouds. 

Our spectroscopic observations were carried out during 2007 October 27$-$29 at
the ESO-VLT (Paranal, Chile) with the high-resolution spectrograph UVES (Dekker
et al. 2000). We used the 0.7" slit (R$\sim$60,000) in the Red Arm Mode
($\sim$5,700$-$9,600 \AA), which gives a resolution of $\sim$0.13 at 7800 \AA\
(the Rb\,{\sc i} line). The exposure times varied between 30 minutes for the
less obscured stars and 3 hours for the Rb-detected stars. The goal was to
achieve a final S/N ratio of $\geq$30$-$50 at 7800 \AA.  All 24 obscured O-rich
AGB candidates (17 in the LMC and 7 in the SMC) were inspected at the telescope
but useful spectra were obtained for only 10. The other 14 sources were either
too faint at 7800 \AA\ or the optical counterpart was not found. In addition, we
observed 10 (4 in the LMC and 6 in the SMC) very luminous and less obscured
stars with previous Li information. For comparison purposes, we also obtained a
spectrum for IRAS 04553$-$6825, a well known LMC massive red supergiant. Table 1
presents relevant information for the sample stars with useful spectra: the
obscured and unobscured O-rich AGBs as well as the massive red supergiant (RSG)
stars. 

The spectra were reduced at the telescope using the UVES data reduction pipeline
(Ballester et al. 2000). Fig. 1 shows sample spectra around the 7800 \AA\
Rb\,{\sc i} line. In general, all stars show extremely red spectra\footnote{Note
that the Rb-detected stars display the reddest spectra with the flux level
falling dramatically at wavelengths shorter than 7000 \AA.}, severely dominated
by strong molecular bands of TiO and sometimes ZrO. Remarkably, the Rb-detected
stars seem to be spectroscopically different from the non Rb-detected stars (see
Table 1) in some spectral regions such as $\sim$7400$-$7600, 7925$-$7975,
8100$-$8150, and 9250$-$9350 \AA\footnote{The non Rb-detected LMC stars RGC 69
and SHV G149006 display spectra identical to the Rb-detected ones and we
identify them as very massive AGBs.}. The latter regions are dominated by
numerous and as yet unidentified molecular features. In addition, the Rb\,{\sc
i} absorption lines may have blue-shifted circumstellar absorption components.
Unfortunately, the accompanying circumstellar Rb\,{\sc i} emission component
sometimes coincides with the photospheric Rb\,{\sc i} absorption. Thus, in some
cases, the circumstellar emission masks the photospheric absorption and this
prevents us measuring a reliable Rb abundance in these stars.

\section{Abundance analysis}

Our abundance analysis combines state-of-the-art line-blanketed model
atmospheres for cool stars and extensive linelists. Basically, we have followed
the  procedure that we used for the galactic O-rich AGB stars (see
Garc\'{\i}a-Hern\'andez et al. 2006, 2007 for more details). The principal
difference is that we have constructed a grid of MARCS model atmospheres
(Gustafsson et al. 2008) at the metallicity of the MCs (we assumed
z=[M/H]=$-$0.3 and $-$0.7 for the LMC and SMC, respectively). Then, we generated
a grid of  MARCS synthetic spectra with effective temperatures in the range
T$_{eff}$=2600$-$4000 K in steps of 100 K, the FWHM in the range 200$-$600 m\AA\
in steps of 50 m\AA\ (this step accounts for the instrumental profile and the
macroturbulence),  and keeping all the other stellar parameters fixed: surface
gravity $log g$=0.0, microturbulence $\xi$=3 km s$^{-1}$, stellar mass  M=1
M$_{\odot}$\footnote{The output synthetic spectra are not sensitive to the mass
of the star in the range 1$-$10 M$_{\odot}$ (Garc\'{\i}a-Hern\'andez et al.
2007).}, and scaled solar abundances.

The observed spectra were compared to the synthetic ones in the region
7775$-$7835 \AA\ that covers the Rb\,{\sc i} line at 7800 \AA; see Fig. 2. We
first determined which of the spectra from our grid of models provides the best
fit to the TiO bandheads and the pseudocontinuum around the Rb\,{\sc i} line by
adjusting mainly the T$_{eff}$. Then, the rubidium content\footnote{The Rb
isotopic and hyperfine structure was considered, for which we assumed a solar Rb
isotopic composition (Garc\'{\i}a-Hern\'andez et al. 2006).} was estimated by
fitting the Rb\,{\sc i} line. When an acceptable S/N ratio ($>$30) was achieved
around 7000 \AA,  the T$_{eff}$ was checked from syntheses of the region
7025$-$7075 \AA\ which includes the TiO red-degraded bandhead at 7054 \AA. We
found a typical difference of $\pm$100 K between the temperatures. In addition,
we derived the Zr abundance or an upper limit from synthesis of the 7440 \AA\
Zr\,{\sc i} line or the ZrO molecular bands in the intervals 6455$-$6499 and
6900$-$6950 \AA (see below). 

A surprising result immediately appeared: the Rb abundances obtained for the
stars not having a strong 7800 \AA\ Rb\,{\sc i} line were unrealistically low 
([Rb/Fe]$\leq$$-$1.7 in some LMC stars and similar limits for SMC stars) given
the anticipated initial Rb abundances for MC stars. It is to be noted that the
metallicity derived from the 7798 \AA\ Ni\,{\sc i} and 7808 \AA\ Fe\,{\sc i}
lines close to the Rb\,{\sc i} line is significantly lower (by up to 1 dex!)
than the one derived from metallic lines (e.g., the Fe\,{\sc i} lines at 7443
and 7446 \AA) around the 7440 \AA\ Zr\,{\sc i} line. For example, in the LMC
Li-rich Rb-low AGB star HV 5584, [Fe/H]$=$$-$1.5 and [Fe/H]$=$$-$0.4 are
obtained from the 7800 \AA\ and 7440 \AA\ regions, respectively. The disparity
in the Fe abundances is likely attributable to a missing opacity in the real
atmosphere and to deficiencies in the adopted TiO line list (see also
Lambert et al. 1995; Abia et al. 2001). The TiO line list plays a much greater
role at 7800 \AA\ than at 7440 \AA. Therefore, we computed the [Rb/z] and [Zr/z]
ratios using the metallicity derived from the metallic lines (calibrated against
the spectrum of Arcturus) close to the 7800 \AA\ Rb\,{\sc i} and 7440 \AA\
Zr\,{\sc i} lines, respectively. However, the analysis of the Rb-detected stars
is an even  trickier business. The available line list for the 7440 \AA\ region
is incomplete for these stars, probably the list is missing lines linked to the
yet unidentified molecule responsible for several bandheads in this region,
bandheads not present in the Rb-weak and less luminous stars. Thus, we are
presently forced to use the ZrO bands to set the Zr abundance in the Rb-detected
stars. At least in the less luminous MC AGB stars such as HV 5584, HV 1645 or HV
11427, the Zr abundance from the 7440 \AA\  Zr\,{\sc i} line is in very good
agreement ($\pm$0.1 dex) with that derived from the ZrO bands.

The spectroscopic effective temperatures and abundances are summarised in Table
1. The uncertainties\footnote{These errors reflect mostly the sensitivity of the
derived abundances to changes in the atmospheric parameters taken for the
modelling.} of the derived abundances are estimated to be 0.8 and 0.5 dex for Rb
and Zr abundances, respectively. The final fit to the 7800 \AA\ region of the
LMC Rb-rich star IRAS 04498-6842 is shown in Fig. 2. 

\section{The low Rb stars}

Setting aside the few Rb-rich stars, non-detection of the 7800 \AA\ Rb\,{\sc i}
line corresponds to a limit [Rb/z]$\leq$$-$0.5 for both AGB and RSG stars.
Obviously, it is of interest to compare this limit with the value for warmer
giants less evolved than AGB and RSG stars and with simpler spectra.
Unfortunately, Rb abundances have not yet been reported for such stars, but Y
and Zr abundances are available. Pomp\'eia et al. (2008) find [Y/Fe] $\simeq
-0.3$ and [Zr/Fe] $\sim -0.5$ for LMC red giants. The limit [Rb/z]$\leq$$-$0.5
is consistent with these values. The AGB stars appear to show a mild Zr excess
over [Zr/Fe] $\sim -0.5$. A mild Rb excess can not yet be ruled out given the
uncertainty affecting the Rb abundance and the possibility that non-LTE effects
have resulted in an underestimate of the Rb abundance (Plez et al. 1993). In
short, the low Rb stars with a mild Zr excess appear to be AGB stars that have
experienced thermal pulses.

\section{The Rb-rich AGBs}

The main result of our survey is that we have discovered strong Rb\,{\sc i}
lines in AGB stars (4 stars in the LMC and 1 in the SMC) in a low-metallicity
extragalactic system. As in our Galaxy, the Rb-strong stars in the MCs belong to
the class of OH/IR stars (Table 1). Unfortunately, we could estimate
photospheric Rb abundances in only three LMC Rb-rich stars due to the clear
presence of blue-shifted circumstellar Rb\,{\sc i} lines in the other
Rb-detected stars (Table 1). The extremely high Rb abundances observed (up to
10$^{3}$$-$10$^{5}$ times solar) among the LMC stars are remarkable. These Rb
abundances at their maximum are a factor of ten or more greater than displayed
by their Galactic counterparts. Note that, when the S/N ratio is high enough at
6708 \AA, as is the case for two of the four LMC Rb-rich stars, the Li\,{\sc i}
6708 \AA\ feature seems to be strong indicating Li-production by a HBB AGB star.
The Li feature is also strong for other AGB stars in both the LMC and SMC that
are not Rb-rich; an indicator that Li synthesis and Rb synthesis are not tightly
coupled. 

The well known period-luminosity relation for luminous AGB stars (Mira
variables) shows that the Rb-rich stars are those with the longest periods. In
addition, a common mark of the Rb-rich stars that sets them apart from other AGB
stars is their bolometric luminosity -- see Fig. 3 for a plot of M$_{bol}$ vs.
[Rb/z]. References to M$_{bol}$ estimates are given in Table 1. Bolometric
magnitude estimates are accurate to about 0.5 magnitudes (Whitelock et al.
2003). The Rb-rich stars in the LMC are brighter ($-$8$<$M$_{bol}$$<$$-$7) than
the Rb-poor stars. Fig. 3 includes the Li-rich HBB-AGBs previously studied in
the SMC (Plez et al. 1993). Unfortunately, we do not have a star in common with
Plez et al. (1993), but our Rb and Zr abundances for the SMC HBB-AGBs agree very
well (within the errors) with their reported abundances. Note, however, that
Plez et al. derived the metallicity from the metallic lines in the 7400$-$7600
\AA\ window which is only slightly blanketed by TiO molecular lines. The use of
the metallic lines near to the Rb\,{\sc i} line will probably bring up their
reported Rb abundances even closer to our values. 

The apparent onset of Rb-rich stars at luminosities of M$_{bol}$ of $-$7.1 is
intriguing. This bolometric luminosity is the generally adopted limit for AGB
stars (Paczy\'nski 1971). Stars more luminous than this limit have been thought
to be massive red supergiants\footnote{Examples in our sample of massive
red supergiants are noted in Table 1. Absence of Li (also Rb and Zr) is a mark
of a RSG (e.g., IRAS 04553-6825)} although presence of AGB stars at luminosities
brighter than the standard limit - as our observations confirm - can be due to a
luminosity contribution from HBB in a massive AGB star. Models
suggest that the Li-rich HBB-AGBs with $-$7$<$M$_{bol}$$<$$-$6 in the LMC are
the descendants of stars with initial masses M $\sim$ 4$-$4.5 M$_{\odot}$
(Ventura et al. 2000): our Rb-rich LMC AGB stars with M$_{bol}$ $<$ $-$7 might
have initial stellar masses of at least $\sim$6$-$7 M$_{\odot}$.

\section{A Rubidium problem}

The Rb problem posed by the four LMC stars has two parts: the high Rb abundance
and the extraordinary [Rb/Zr] ratio (i.e., the apparent lack of a Zr
enrichment). Our discovery of the class of Rb-rich LMC AGB stars is assured by
visual inspection of our spectra (Fig. 1). The severity of the Rb overabundance
([Rb/z]$ \sim  +2.8$ to $+5.0$) may be somewhat uncertain because the Rb\,{\sc
i} line is strong and saturated with possible circumstellar contamination.  Note
that all Rb-detected stars also display strong Rb\,{\sc i} lines at 7947 \AA,
confirming the high Rb abundance from the 7800 \AA\ line. For example,
[Rb/z]=$+$5.0 and $+$3.3 are obtained from the 7947 \AA\ Rb\,{\sc i} line for
the Rb-rich AGBs IRAS 04498$-$6842 and IRAS 04407$-$7000, respectively. The
upper limit to the Zr abundance that gives ratios [Rb/Zr] of $\>$3 to 4 (Table
1) comes from a fit to ZrO bands.

A Rb overabundance is naturally attributed to the $s$-process and most probably
to its operation in massive AGB stars with the higher neutron density from the
$^{22}$Ne source taking the $s$-process path through the $^{85}$Kr branch to
$^{87}$Rb and resulting in an increase in Rb abundance relative to the low
neutron density path to $^{85}$Rb (Garc\'{\i}a-Hern\'{a}ndez et al. 2006).
Although the increase in the Rb abundance between low and high neutron density
$s$-process paths is about an order of magnitude, the predicted Rb/Zr and Rb/Y
ratios do not assume extreme values. Present massive AGB nucleosynthesis models
can qualitatively describe the observations of Rb-rich AGBs in the sense that
increasing Rb overabundances with increasing stellar mass and with decreasing
metallicity are theoretically predicted (van Raai et al. 2008, 2009). However,
these theoretical models are far from matching the extremely high Rb
enhancements that we observe. Predictions for massive AGB models at the LMC and
SMC metallicities (and solar for comparison) computed for the $^{22}$Ne
source with the Monash stellar nucleosynthesis code based on the Mt. Stromlo
stellar structure code (e.g., Karakas \& Lattanzio 2007) are shown in Table 2.
If the ``3$^{rd}$ dredge-up" efficiency remains as high as before the onset of
the superwind phase during the final few pulses of a massive AGB star, then
[Rb/Fe] increases as well as [Zr/Fe] (e.g., up to $+$1.3 and $+$0.8,
respectively, in the M=6 M$_{\odot}$, LMC case). Even considering higher AGB
masses, within the framework of the s-process it is not possible to produce
extremely high Rb abundances without co-producing Zr at similar levels because
both Rb and Zr belong to the first s-process peak.

The extraordinary [Rb/Zr] values are likely artefacts of the analysis and
possibly a result of the necessity of using the ZrO bands to set the Zr
abundance. Non-LTE effects and a failure of the adopted models to represent the
real stars are surely contributing factors too. If the large [Rb/Zr] values are
real, we can offer no explanation in terms of nucleosynthesis. Additional
observations of luminous AGB stars are needed to confirm that Rb-rich stars are
confined to bolometric magnitudes M$_{bol} = -7.1$ and brighter, and that the
SMC also contains similar stars to the four LMC examples. Despite the
uncertainties in the Rb abundance determinations, the occurrence of Rb-rich
stars among the most luminous AGB stars -- HBB stars as indicated by the
presence of Li -- in the LMC is assured. 



\acknowledgments

This work is based on observations made at ESO, 080.D-0508(A). We are very
grateful to R. Gallino and L. Siess for helpful discussions. DLL wishes to thank
the Robert A. Welch Foundation of Houston, Texas for support through grant
F-634.



{\it Facilities:} \facility{VLT:Kueyen (UVES)}.

\clearpage

\begin{deluxetable}{lcccccccc}
\tabletypesize{\scriptsize}
\tablecaption{The MC O-rich AGB sample: spectroscopic temperatures and abundances$^{a}$ \label{tbl-1}}
\tablewidth{0pt}
\tablehead{
\colhead{Name} &   \colhead{T$_{eff}$} & \colhead{[Rb/z]$^{b}$} &\colhead{[Zr/z]$^{c}$} & \colhead{Li\,{\sc i}$^{d}$} &
\colhead{M$_{bol}$} &
\colhead{Period} &  \colhead{Ref$^{e}$.} &\colhead{Type$^{f}$}\\
}
\startdata
 &   &  &  & LMC &  &  &  &  \\
\hline
IRAS 04498$-$6842 & 3400 & $+$5.0	     & $\leq$$+$0.3$^{*}$& yes    & $-$7.72 & 1292 & 1   & OH/IR      \\ 
IRAS 04407$-$7000 & 3000 & $+$3.2	     & $\dots$  	 & $\dots$& $-$7.11 & 1199 & 1   & OH/IR       \\
IRAS 04516$-$6902 & 3000 & $+$3.2$^{*}$      & $\leq$$+$0.3$^{*}$& yes    & $-$7.11 & 1091 & 1   & OH/IR?      \\ 
IRAS 05558$-$7000 & 3400 & $+$2.8	     & $\dots$  	 & $\dots$& $-$6.97 & 1220 & 1   & OH/IR       \\
IRAS 05298$-$6957 & 4000 &$\leq$$-$0.3$^{*}$ & $+$0.0		 & no	  & $-$6.72 & 1280 & 1   & OH/IR       \\
IRAS 05329$-$6708 & 3900 &  $-$0.5	     & $+$0.5		 & no	  & $-$6.95 & 1262 & 1   & OH/IR       \\
IRAS 04553$-$6825 & 3400 & $\leq$$-$0.5      & $-$0.5$^{*}$	 & no	  & $-$9.19 &  841 & 1   & RSG         \\
RGC 15       	  & 3400 & $+$0.1	     & $\leq$$-$0.2$^{*}$& no	  & $-$6.53 &  760 & 2   & non HBB-AGB  \\
HV 5584      	  & 3500 & $\leq$$-$0.5      & $+$0.1		 & yes    & $-$6.27 &  500 & 3   & HBB-AGB	\\
RGC 69       	  & 3300 & $\leq$$-$0.5      & $-$0.2$^{*}$	 & yes    & $-$6.49 &  648 & 2   & HBB-AGB	\\
SHV G149006  	  & 3100 & $\leq$$-$0.7      & $-$0.2$^{*}$	 & yes    & $-$6.65 &  636 & 4   & HBB-AGB	\\
\hline
 &   &  &  & SMC &  &  &  &    \\
\hline
IRAS 00483$-$7347 & 3400 & $+$1.7$^{*}$  & $\dots$	     &$\dots$& $-$7.20 & 1200	     & 5		& OH/IR?     \\
IRAS 00591$-$7307 & 3400 & $\leq$$-$1.0  & $-$0.2	     & no    & $-$8.30 & $\geq$1000  & 6		&RSG/AGB?    \\ 
IRAS 01082$-$7335 & 3300 & $\leq$$-$1.0  & $\leq$$+$0.2$^{*}$& no    & $\dots$&$\dots$       & $\dots$  	&RSG	\\
MSX SMC 168       & 3700 & $-$1.0	 & $-$0.1	     & no    & $\dots$&$\dots$       & $\dots$  	&RSG	 \\
HV 838            & 3400 & $\leq$$-$0.5  & $+$0.2	     & yes   & $-$7.18 &  663        & 3 	       & HBB-AGB    \\
HV 1645           & 3400 & $\leq$$-$0.5  & $-$0.4	     & yes   & $-$4.68 &  300        & 3 	       & HBB-AGB    \\
HV 1719           & 3400 & $+$0.4$^{*}$  & $+$0.7$^{*}$      & yes   & $-$6.68 &  531        & 3 	       & HBB-AGB    \\
HV 11427          & 3500 & $\leq$$-$0.5  & $+$0.0	     & no    & $-$5.04 &  251        & 3 	       & non HBB-AGB  \\
NGC371 R20        & 3900 & $-$0.5	 & $-$0.2	     & no    & $-$8.36 &  580        & 3 	       & RSG	     \\
NGC371 29         & 3400 & $\leq$$+$0.0  & $-$0.3	     & no    & $-$7.36 &  530        & 3 	       & RSG/AGB?    \\
\enddata
\tablenotetext{a}{Relative to the solar photospheric values: log$\varepsilon$(Rb, Zr)=2.6.} 
\tablenotetext{b}{For T$_{eff}$ $\geq$ 3300 K, z is the average of the Fe and Ni
abundances estimated from the closest Fe \,{\sc i} and Ni \,{\sc i} lines, while for
T$_{eff}$ $<$ 3300, z=[M/H]=$-$0.3 and $-$0.7 are assumed for the LMC and SMC,
respectively. The asterisk means that the Rb\,{\sc i} line has a circumstellar origin
and the given value corresponds to the photospheric abundance needed to fit the depth
of the observed circumstellar line. These abundances must be considered with caution
but they usually represent a rough estimation of the photospheric Rb content.}
\tablenotetext{c}{Zr abundances from the 7440 \AA\ Zr\,{\sc i} line where z is
the Fe abundance estimated from the closest Fe \,{\sc i} lines. The asterisk
means that the Zr abundance is estimated from the ZrO molecular bands where,
because of the lack of useful metallic lines, z=[M/H]=$-$0.3 and $-$0.7 are
assumed for the LMC and SMC, respectively. Note that no entry means that the S/N
is too low for Zr abundance determinations.}
\tablenotetext{d}{No entry means that the S/N is too low to infer the presence of
Li\,{\sc i} at 6708 $\AA$. The detection or non-detection of Li generally gives a lower
and upper limit of log$\varepsilon$(Li)=12$+$logN(Li)$>$1.0 and $<$0.0, respectively.}
\tablenotetext{e}{References for bolometric magnitude and period of variability.}
\tablenotetext{f}{OH/IR is assigned to very luminous stars (M$_{bol}$ $\leq$
$-$6.7) showing OH maser emission, extremely long periods ($>$ 1000 days) and
large amplitude of variability ($\Delta$K$>$1.2 mag), being very obscured
(J$-$K$\geq$3) and very bright in the mid-infrared (F$_{25}$ $\geq$1 Jy, when
detected by the IRAS satellite). In contrast, massive Red Supergiant (RSG) stars
are even more luminous than OH/IRs and they are characterized by a small
amplitude of variability ($\Delta$K$<$0.5 mag). HBB-AGBs were previously known to
be Li-rich (Smith et al. 1995) and they generally are less luminous and obscured
than OH/IRs and they are characterized by periods shorter than 700 days.} 
\tablerefs{(1) Whitelock et al. (2003); (2) Reid, Glass \& Catchpole (1988); (3)
Wood, Bessel \& Fox (1983); (4) Hughes \& Wood (1990); (5) Whitelock et al.
(1989); (6) Elias, Frogel and Humphreys (1980,1985).}
\end{deluxetable}

\begin{deluxetable}{lccc}
\tabletypesize{\scriptsize}
\tablecaption{Predictions for massive AGB models.  \label{tbl-2}}
\tablewidth{0pt}
\tablehead{
\colhead{Mass} &   \colhead{z} & \colhead{[Zr/Fe]} & \colhead{[Rb/Fe]}
}
\startdata
5 M$_{\odot}$ & solar  &  $+$0.01  &  $+$0.05   \\
6 M$_{\odot}$ & solar  &  $+$0.07  &  $+$0.21   \\
5 M$_{\odot}$ & LMC    &  $+$0.21  &  $+$0.56   \\
6 M$_{\odot}$ & LMC    &  $+$0.41  &  $+$0.79   \\
5 M$_{\odot}$ & SMC    &  $+$1.07  &  $+$1.11   \\
\enddata
\end{deluxetable}

 \clearpage
 

\begin{figure}
\includegraphics[angle=-90,scale=.60]{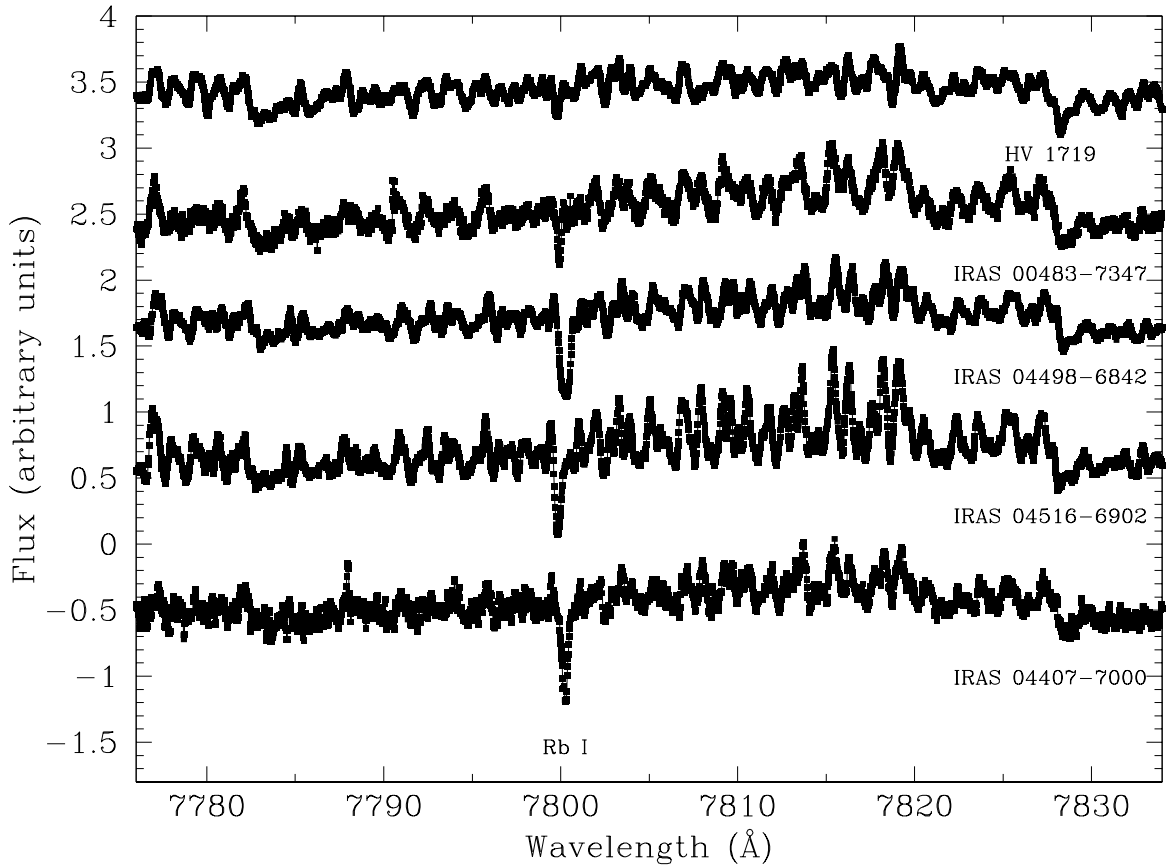}
\caption{Sample spectra around the 7800\AA\ Rb\,{\sc i} line.  The lower three
stars (IRAS 00498-6842, 04516-6902, and 04407-7000) are Rb-rich LMC stars. The
upper two stars (HV 1719 and IRAS 00483-7347) with a much weaker Rb\,{\sc i} line
are from the SMC.
\label{fig1}}
\end{figure}

\clearpage

\begin{figure}
\includegraphics[angle=-90,scale=.60]{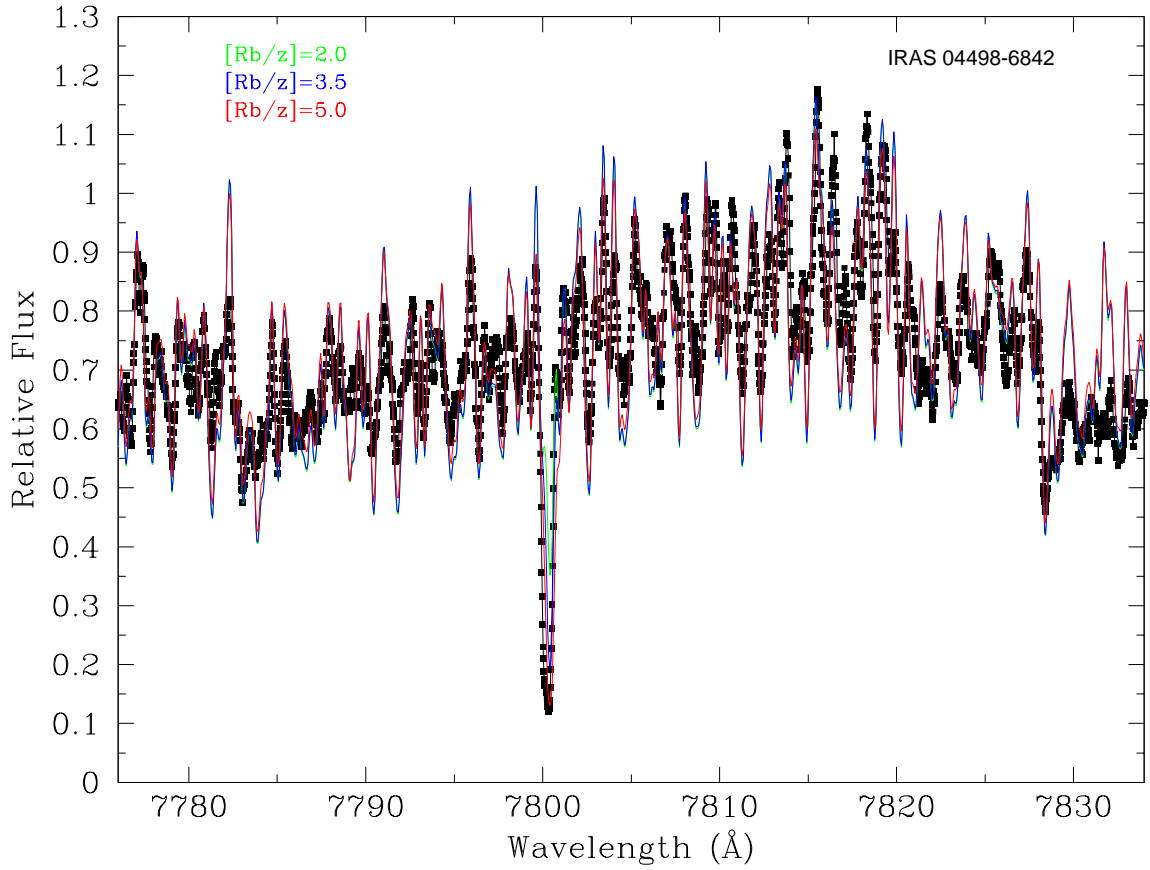}
\caption{Fit of the best synthetic spectrum (red) to the observed spectrum
(black) around the 7800\AA\ Rb\,{\sc i} line for the Rb-rich LMC star IRAS
04498$-$6842. The synthetic spectra obtained for [Rb/z]=$+$2.0 (green) and
[Rb/z]=$+$3.5 (blue) are also shown.
\label{fig2}}
\end{figure}

\clearpage

\begin{figure}
\includegraphics[angle=-90,scale=.60]{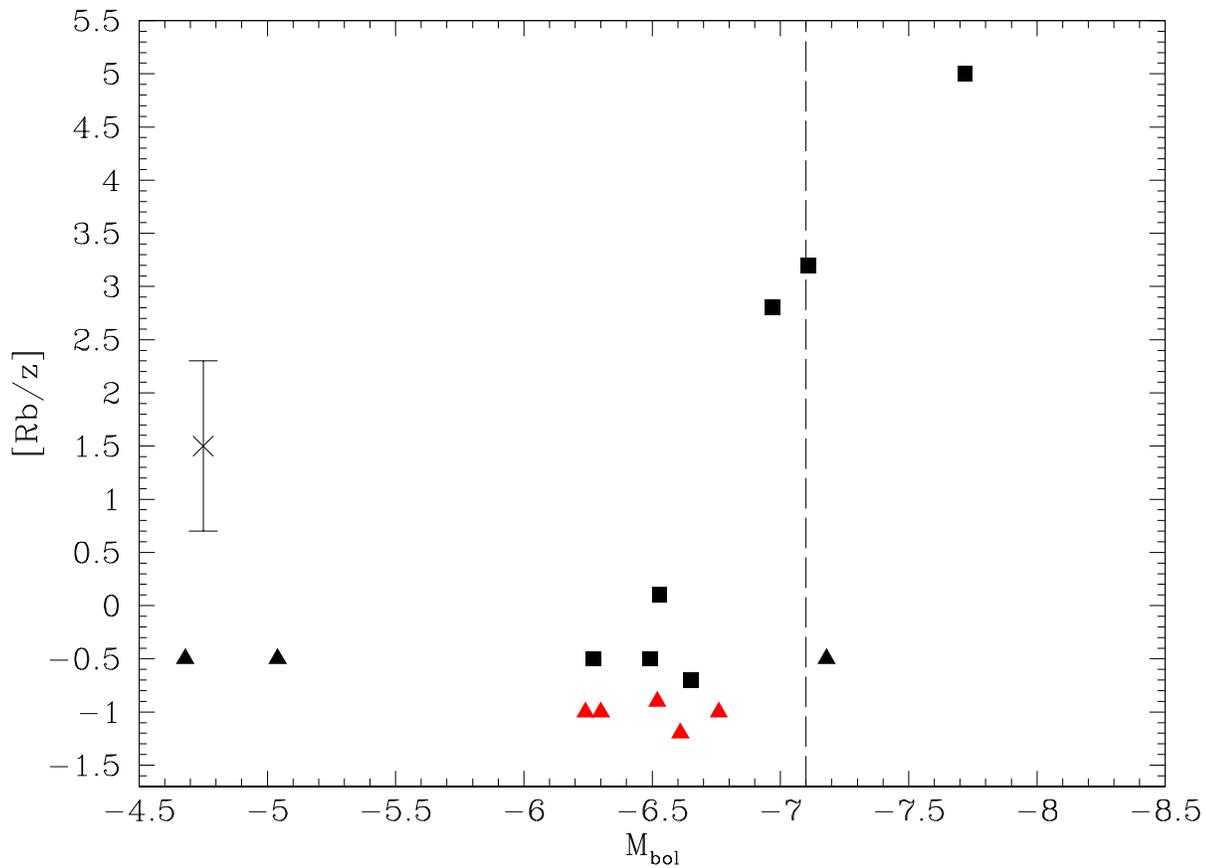}
\caption{Observed Rb abundances (squares and triangles for the LMC and SMC,
respectively) versus M$_{bol}$. Red triangles correspond to the visually bright
HBB-AGB SMC stars previously studied by Plez et al. (1993). A typical error bar
of $\pm$0.8 dex is shown. The dashed vertical line marks the theoretical
luminosity limit (M$_{bol}$=$-$7.1; Paczy\'nski 1971) for AGB stars.
\label{fig3}}
\end{figure}




\end{document}